\documentclass[a4paper]{jpconf}
\usepackage{graphicx}
\begin{document}
\title{Modelling the current accelerated expansion of the Universe with Holographic Dark Energy}

\author{Carlos Rodriguez-Benites$^1$, Mauricio Cataldo$^1$, Antonella Cid$^1$ and Carlos R\'ios$^2$}

\address{$^1$ Departamento de F\'isica, Grupo Cosmolog\'ia y Part\'iculas Elementales, Universidad del B\'io-B\'io, Casilla 5-C, Concepci\'on, Chile.\\
$^2$ Departamento de Ense\~nanza de las Ciencias B\'asicas, Universidad Cat\'olica del Norte, Larrondo 1281, Coquimbo, Chile.}

\ead{crodriguez@ubiobio.cl}

\begin{abstract}
In this work we explore a Holographic Dark Energy Model in a flat Friedmann-Lema\^itre-Robertson-Walker Universe, which contains baryons, radiation, cold dark matter and dark energy within the framework of General Relativity. Furthermore, we consider three types of phenomenological interactions in the dark sector. With the proposed model we obtained the algebraic expressions for the cosmological parameters of our interest: the deceleration and coincidence parameters. Likewise, we graphically compare the proposed model with the $\Lambda$CDM model.
\end{abstract}

\section{Introduction}

Nowadays it is well known that cosmological models must describe an accelerated expansion of the Universe at the present era~\cite{Riess, Perlmutter:1998np, Planck}. To achieve this, sources of matter capable of generating this acceleration are considered, which are commonly dubbed dark energy~\cite{DarkE}.

A cosmological constant $\Lambda$ is an important candidate for dark energy providing a good explanation for the current acceleration. But the cosmological constant faces some problems~\cite{Copeland, Riess2} such as, the mismatch between the expected value of the vacuum energy density and the energy density of the cosmological constant, and the lock of an explanation of why densities of dark energy and dark matter are of same order at present while they evolve in rather different ways. So, as an alternative, dynamical dark energy models have been proposed and analyzed in the literature. Among these, Holographic Dark Energy Models~\cite{Granda, Gao, SergiodelCampo, Lepe, Fabiola, chimentointeracciongeneral} are interacting, because the originate from the holographic principle~\cite{Maldacena}. The holographic principle asserts that the number of relevant degrees of freedom of a system dominated by gravity must vary along with the area of the surface bounding the system~\cite{Hooft}. According to this principle, the vacuum energy density can be bounded~\cite{Cohen} as $\rho_x\leq M^2_pL^{-2}$, where $\rho_x$ is the dark energy density (the vacuum energy density), $M_p$ is the reduced Planck mass, and $L$ is the size of the region (i.e IR cutoff). This bound implies that, the total energy inside a region of size $L$, should not exceed the mass of a black hole of the same size. From effective quantum field theory, an effective IR cutoff can saturate the length scale, so that the dark energy density  can be written as $\rho_x=3c^2M^2_pL^{-2}$~\cite{Li}, where $c$ is a dimensionless parameter, and the factor $3$ is for mathematical convenience. In the Holographic Ricci Dark Energy Model, $L$ is given by the average radius of the Ricci scalar curvature $|\mathcal{R}|^{-1/2}$, so in this case the density of the Holographic Dark Energy (hereafter, abbreviated as HDE) is $\rho_x\propto\mathcal{R}$.
 
In a spatially flat universe, the Ricci scalar of the spacetime is given by $\mathcal{R} =6(\dot{H} + 2H^2)$, where $H(t) =\dot{a}(t)/a(t)$ is the Hubble expansion rate of the universe in terms of the scale factor $a$, where the dot denotes the derivative with respect to the cosmic time $t$. In this sense, the authors of reference~\cite{Granda} introduced the following generalization:
\begin{equation}\label{EOTH}
\rho_x=3(\alpha H^2 + \beta \dot{H})\,,
\end{equation}
where $\alpha$ and $\beta$ are constants to be determined. This  model  works  fairly  well  in  fitting  the  observational  data,  and it is a good candidate to alleviate the cosmic coincidence  problem~\cite{Gao, SergiodelCampo, Lepe, Fabiola, Mathew}.

\section{Basic Equations}

In the framework of General Relativity we consider a homogeneous, isotropic and flat universe scenario through the Friedmann-Lema\^itre-Robertson-Walker (FLRW) metric~\cite{Ryden}
\begin{equation}
ds^2=dt^2-a^2(t)[dr^2+r^2(d\theta^2+\sin^2\theta d\phi^2)] \,,
\end{equation}
where $(t,r,\theta,\phi)$ are comoving coordinates. Friedmann's equations in this context are written as
 \begin{eqnarray}\label{EEH1}
 3\, H^2 &=& \rho\, ,\\
 \label{EEH2}
 2\, \dot{H} + 3\, H^2 &=& -\, p \, ,
 \end{eqnarray}
where $\rho$ is the total energy density, $p$ is the total pressure and  $8\pi G=c=1$ is assumed.
On the other hand, the conservation of the total energy-momentum tensor is given by~\cite{Ryden}
 \begin{equation}\label{EcuaciondeConservacion}
\dot{\rho}+3H(\rho+p)=0\,.
 \end{equation}

\section{Holographic Dark Energy Model}

We study a scenario that contains baryons, radiation, cold dark matter and HDE, i.e. $\rho= \rho_b+\rho_r+\rho_c+\rho_x$ and $p= p_b+p_r+p_c+p_x$. In addition, we consider a barotropic equation of state for the fluids, $p_i=\omega_i\,\rho_i$ with $\omega_b=0$, $\omega_r=1/3$, $\omega_c=0$ and $\omega_x=\omega$. By including a phenomenological interaction in the dark sector, we split the conservation equation~(\ref{EcuaciondeConservacion}) in the following equations
\begin{equation}\label{4FluidosEC3'}
\rho'_c+\rho_c =-\Gamma\quad \textrm{and} \quad \rho'_x+(1+\omega)\,\rho_x =\Gamma\,,
\end{equation}
where prime denotes a derivative with respect to $\ln a^3$ and $\Gamma$ represents the interaction function between cold dark matter and the HDE. From Eqs.~(\ref{EOTH}) and~(\ref{EEH1}) we obtain  
\begin{equation}\label{4FluidosEOTH}
\rho_x=\alpha\,\rho+\frac{3\beta}{2}\,\rho'\,.
\end{equation}

Given that radiation and baryons are separately conserved, we have $\rho_r\propto a^{-4}$ and $\rho_b\propto a^{-3}$. From here it is easy to realize that $\rho''_b=-\rho'_b=\rho_b$ and $\rho''_r=-\frac{4}{3}\rho'_r=\frac{16}{9}\rho_r$.
 
On the other hand, in the study of HDE scenarios usually it is only considered the dark sector, since these predominate in the current universe. Also, it is possible to analyze a HDE scenario with two different approaches, the first one considers a variable state parameter for the HDE or assuming a parameterization as shown in~\cite{Fabiola}, while the second approach considers an interaction term between the dark components~\cite{Gao, chimentointeracciongeneral, interaccionconterminocomomateria}. We work in the last approach.

For convenience, we denote the energy density of the dark sector as $\rho_d:=\rho_c+\rho_x$. Then, by combining equations~(\ref{4FluidosEC3'}) -~(\ref{4FluidosEOTH}) we obtain
\begin{equation}\label{4FluidosEDRho}
\frac{3\beta}{2}\,\rho''_d+\left( \alpha+\frac{3\beta}{2}-1 \right)\,\rho'_d+(\alpha-1)\,\rho_d+\frac{1}{3}(2\beta-\alpha)\,\rho_{r0}\,a^{-4}=\Gamma\,,
\end{equation}	
where the submipt $0$ denotes a current value. Notice that the eq.~(\ref{4FluidosEDRho}) can be easily solve when $\Gamma=\Gamma(\rho_d,\rho'_d,\rho,\rho')$.
In this work we consider the following types of linear interactions~\cite{modeloscosmologicosCataldo,AICprofAntonella, Bayesian}: $\Gamma_1=\alpha_1\rho_c+\beta_1\rho_x$, $\Gamma_2=\alpha_2\rho'_c+\beta_2\rho'_x$, and $\Gamma_3=\alpha_3\rho_d+\beta_3\rho'_d$.

\subsection{The energy density of the dark sector}

We can convenient rewrite eq.~(\ref{4FluidosEDRho}) as
\begin{equation}\label{4FluidosEDRhoGeneral}
\rho''_d+b_1\,\rho'_d+b_2\,\rho_d+b_3\,a^{-3}+b_4\,a^{-4}=0\,,
\end{equation}
including the three interaction types of our interest where the values of the constants $b_1$, $b_2$, $b_3$ and $b_4$ are shown in Table~\ref{tabla:sencilla}. The general solution of eq.~(\ref{4FluidosEDRhoGeneral}) is:
\begin{equation}\label{4FluidosEDRhoSolGeneral}
\rho_d(a)=A\,a^{-3} +B\,a^{-4}+C_1\,a^{3\,\lambda_1}+C_2\,a^{3\,\lambda_2}\,,
\end{equation}
where the integration constants $C_1$ and $C_2$ are given by
\begin{eqnarray}
\nonumber C_1&=& \frac{3A\beta(1+\lambda_2)+B\beta(4+3\lambda_2)+3H_0^2\,(-2\alpha+2\Omega_{x0}+\beta(3\Omega_{b0}+4\Omega_{r0}-3\lambda_2(\Omega_{c0}+\Omega_{x0})))}{3\beta(\lambda_1-\lambda_2)}\,,\\
C_2&=&-A-B+3H_0^2(\Omega_{c0}+\Omega_{x0})-C_1\,,
\end{eqnarray}
where $H_0$, $\Omega_{c0}$, and $\Omega_{x0}$ are the current values of the Hubble parameter, the density parameters for dark matter and HDE (i.e. $\Omega_{i0}=\rho_{i0}/3H^2_0$ with $i=\{c,x\}$), respectively. The coefficients in eq.~(\ref{4FluidosEDRhoSolGeneral}) are $A = \frac{b_3}{b_1-b_2-1}$ and $B = \frac{9b_4}{12b_1-9b_2-16}$, as well as $\lambda_{1,2}=-\frac{1}{2}\left(b_1\pm\sqrt{b^2_1-4b_2}\right)$.

\begin{table}[htbp]
	\begin{center}
		\caption{Definition of the constants $b_1$, $b_2$, $b_3$ and $b_4$ in terms of the model's parameters for the studied interactions.}
		\vspace{0,3cm}
			\begin{tabular}{| c | c | c | c |}
			\hline
			& $\Gamma_1=\alpha_1\,\rho_c+\beta_1\,\rho_x$ & $\Gamma_2=\alpha_2\,\rho'_c+\beta_2\,\rho'_x$ & $\Gamma_3=\alpha_3\,\rho_d+\beta_3\,\rho'_d$  \\
			\hline 
			\hline
			$b_1$ & \scriptsize{$1+\alpha_1-\beta_1-\frac{2}{3\beta}(1-\alpha)$} & \scriptsize{ $\frac{2\alpha-3\beta-2-2\alpha_2-2\alpha(\beta_2-\alpha_2)}{3\beta(1-\beta_2+\alpha_2)}$} & \scriptsize{$\frac{2}{3\beta}\left( \alpha+\frac{3\beta}{2}-1-\beta_3 \right)$}\\
			\hline
			$b_2$ & \scriptsize{$\frac{2}{3\beta} (\alpha(1-\beta_1+\alpha_1)-1-\alpha_1)$ }& \scriptsize{$\frac{2(\alpha-1)}{3\beta(1-\beta_2+\alpha_2)}$ }& \scriptsize{$\frac{2}{3\beta}(\alpha-1-\alpha_3)$} \\ 
			\hline
			$b_3$ & \scriptsize{$(\beta_1-\alpha_1)\left( 1-\frac{2\alpha}{3\beta}  \right) \rho_{b0}$} &\scriptsize{ $\frac{(2\alpha-3\beta)(\beta_2-\alpha_2)}{3\beta(1-\beta_2+\alpha_2)}\rho_{b0}$} &\scriptsize{ $0$}  \\ 
			\hline
			$b_4$ &{\scriptsize{ $\frac{2}{3\beta}\left(\frac{1}{3}(2\beta-\alpha)-(\beta_1-\alpha_1)(\alpha-2\beta) \right) \rho_{r0}$}} & \scriptsize{$\frac{2(2\beta-\alpha)-8(2\beta-\alpha)(\beta_2-\alpha_2)}{9\beta(1-\beta_2+\alpha_2)}\rho_{r0}$} & \scriptsize{$\frac{2}{9\beta}(2\beta-\alpha)\rho_{r0}$} \\ 
		 \hline
		\end{tabular}
		\label{tabla:sencilla}
	\end{center}
\end{table}

\subsection{The state parameter of the HDE}

The state parameter of the HDE corresponds to the ratio $\omega=\frac{p_x}{\rho_x}$. Using the expression~(\ref{4FluidosEOTH}) in eq.~(\ref{4FluidosEC3'}), and the linear interactions $\Gamma_i$, we find
\begin{equation}\label{4FluidosWcomofunciondea}
\omega(a)=\frac{D_1\, a^{-3} +D_2\, a^{-4}+D_3\,a^{3\,\lambda_1}+D_4\,a^{3\lambda_2}}{\tilde{A}\, a^{-3} + \tilde{B}
	\, a^{-4}+\tilde{C}_1\,a^{3\,\lambda_1}+\tilde{C}_2\,a^{3\lambda_2}}\,, 
\end{equation}
where $\tilde{A}=(2\alpha-3\beta)(A+\rho_{b_0})$, $\tilde{B}=2(\alpha-2\beta)(B+\rho_{r_0})$, $\tilde{C}_{1,2}=C_{1,2}(3\beta\lambda_{1,2}+2\alpha)$ and the constant coefficients $D_i$ are shown in table~\ref{tabla:W(a)}.

\begin{table}[htbp]
	\begin{center}
		\caption{Definition of the constants $D_1$, $D_2$, $D_3$ and $D_4$ in terms of the model's parameters for the studied interactions.}
		\vspace{0,3cm}
		\begin{tabular}{| c | c | c | c |}
			\hline
			&\scriptsize{$\Gamma_1=\alpha_1\,\rho_c+\beta_1\,\rho_x$} &\scriptsize{$\Gamma_2=\alpha_2\,\rho'_c+\beta_2\,\rho'_x$ }& \scriptsize{$\Gamma_3=\alpha_3\,\rho_d+\beta_3\,\rho'_d$}  \\
			\hline 
			\hline
			\scriptsize{$D_1$} & \tiny{$2\alpha_1A+(2\alpha-3\beta)(\beta_1-\alpha_1)(A+\rho_{b0})$} & \tiny{ $-2\alpha_2A+(3\beta-2\alpha)(\beta_2-\alpha_2)(A+\rho_{b0})$} & \tiny{$2(\alpha_3+\beta_3)A$} \\
			\hline
			\scriptsize{$D_2$} & \tiny{$2\alpha_1B+2(\alpha-2\beta)\left(\frac{1}{3}-\alpha_1+\beta_1\right)(B+\rho_{r0})$ }& \tiny{$-\frac{8}{3}\alpha_2B+\frac{2}{3}(2\beta-\alpha)(-1-\alpha_2+\beta_2)(B+\rho_{r0})$ }& \tiny{$2\left(\alpha_3-\frac{4}{3}\beta_3\right)B+\frac{2}{3}(\alpha-2\beta)(B+\rho_{r0})$} \\ 
			\hline
			\scriptsize{$D_3$} & \tiny{$C_1(2\alpha_1+(2\alpha+3\beta\lambda_1)(\beta_1-\alpha_1-1-\lambda_1))$} &\tiny{$C_1(2\alpha_2\lambda_1-(2\alpha+3\beta\lambda_1)(1+\lambda_1(1+\alpha_2-\beta_2)))$} &\tiny{$C_1(2(\alpha_3+\beta_3\lambda_1)-(2\alpha+3\beta\lambda_1)(1+\lambda_1))$}  \\ \hline 
			\scriptsize{$D_4$} &{\tiny{ $C_2(2\alpha_1+(2\alpha+3\beta\lambda_2)(\beta_1-\alpha_1-1-\lambda_2))$}} & \tiny{$C_2(2\alpha_2\lambda_2-(2\alpha+3\beta\lambda_2)(1+\lambda_2(1+\alpha_2-\beta_2)))$} & \tiny{$C_2(2(\alpha_3+\beta_3\lambda_2)-(2\alpha+3\beta\lambda_2)(1+\lambda_2))$} \\ 
			\hline
		\end{tabular}
		\label{tabla:W(a)}
	\end{center}
\end{table}

In the limit to the future, $a\rightarrow \infty$, the expression~(\ref{4FluidosWcomofunciondea}) remains as $\omega=\frac{D_3}{C_1(3\beta\lambda_1+2\alpha)}$ for $\lambda_1> \lambda_2>0$, while for $\lambda_2> \lambda_1>0$, we have $\omega=\frac{D_4}{C_2(3\beta\lambda_2+2\alpha)}$.

\subsection{The coincidence and deceleration parameters}

To examine the problem of cosmological coincidence, we define $r\equiv\rho_c/\rho_x$. Then, using $\rho_c=\rho_d-\rho_x$, together with the expression~(\ref{4FluidosEOTH}), we find
\begin{eqnarray}\label{Solder}
r=\frac{\rho_d}{\left(\alpha-\frac{3\beta}{2}\right)\rho_b+(\alpha-2\beta)\rho_r+\alpha\rho_d+\frac{3\beta}{2}\rho'_d}-1\,.
\end{eqnarray}

Then, for all our interactions we get $r(a\rightarrow \infty)=\frac{2}{2\alpha+3\beta\lambda_i}-1$, a constant that depends on the interaction parameters, where $\lambda_i=\textrm{max}\{\lambda_1,\lambda_2\}$ for $\lambda_i>0$.

On the other hand, the deceleration parameter $q$ is a dimensionless measure of the cosmic acceleration in the evolution of the universe. It is defined by $q\equiv -\left(1+\frac{\dot{H}}{H^2} \right)=-\left(1+\frac{3\rho'}{2\rho}\right)$~\cite{Ryden}. Using~(\ref{4FluidosEDRhoSolGeneral}), we obtain
\begin{equation}
q(a)=-\left( 1+\frac{-3(\rho_{b0}+A)a^{-3}-4(\rho_{r0}+B)a^{-4}+3(C_1\lambda_1a^{3\lambda_1}+C_2\lambda_2a^{3\lambda_2})}{2(\rho_{b0}+A)a^{-3}+2(\rho_{r0}+B)a^{-4}+2(C_1a^{3\lambda_1}+C_2a^{3\lambda_2})} \right)\,.
\end{equation}

\begin{figure}[htbp]
\centering\includegraphics[scale=0.35]{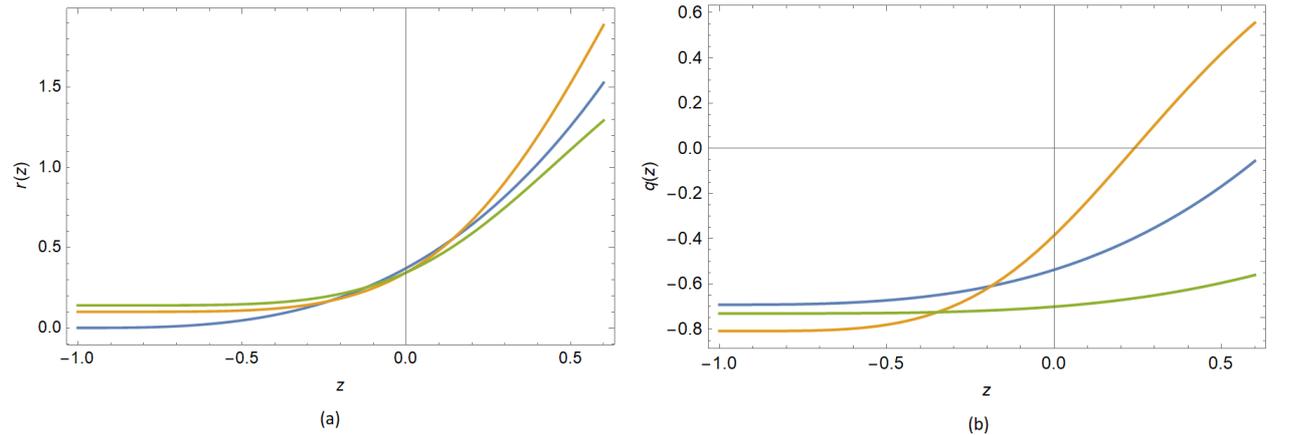}
\caption{(a) Evolution of coincidence parameter $r$ as a function of redshift $z$. (b) Evolution of deceleration parameter $q$ as a function of redshift $z$. In the figures, $z=0$ represents current time.}\label{r(z)}
\end{figure}
In Fig.~\ref{r(z)} we show the evolution of the coincidence and deceleration parameters in term of the redshift $z$, where $a(z)=(1+z)^{-1}$. We use the following values for the parameters~\cite{Planck}: $\Omega_{b0}=0.0484$, $\Omega_{r0}=1.25\times 10^{-3}$, $\Omega_{c0}=0.258$, $\Omega_{x0}=0.692$,
$H_0=67.8$ km s$^{-1}$ Mpc$^{-1}$, and $\omega_{\Lambda CDM}=-1$. In addition, $(\alpha_1, \beta_1)=(-0.0076, 0)$ and $(\alpha_2, \beta_2)=(0.0074, 0)$~\cite{AICprofAntonella, Bayesian} are considered. The blue line represents $\Lambda$CDM, the orange line the model $\Gamma_1$ with $(\alpha,\beta)=(0.86,0.46)$ and the green line the model $\Gamma_2$ with $(\alpha,\beta)=(1.01,0.45)$. We note that the HDE models resemble the $\Lambda$CDM model, in the evolution of both parameters. Besides, in the cases shown for the HDE models with interaction $\Gamma_1$ and $\Gamma_2$, the problem of cosmological coincidence is alleviated, given that the coincidence parameter $r$ tend asymptotically to a positive constant.

\section{Final Remarks}

A theoretical model was developed according to the current components of the Universe, such as baryons, radiation, cold dark dark and HDE, with interaction in the dark sector, obtaining for the HDE, the functions $\omega(z)$, $r(z)$ and $q(z)$. The proposed model was compared graphically with $\Lambda$CDM, using referential values for the HDE parameters and the given interactions.

In the near future we expect to contrast the present scenarios with the observational data (SNe Ia, CC, BAO, CMB), using Bayesian statistics. 

\section*{Acknowledgments}

This research was supported by Universidad del B\'io-B\'io through ``Beca de Postgrado'', and research projects DIUBB 181907 4/R (MC) and GI 172309/C (MC, AC).


\section*{References}
\begin{thebibliography}{9}

\bibitem{Riess}
    A.~G.~Riess {\it et al.} [Supernova Search Team],
  Astron.\ J.\  {\bf 116}, 1009 (1998)
  [astro-ph/9805201]. 

\bibitem{Perlmutter:1998np} 
  S.~Perlmutter {\it et al.} [Supernova Cosmology Project Collaboration],
  Astrophys.\ J.\  {\bf 517}, 565 (1999)
  [astro-ph/9812133].

\bibitem{Planck} 
  P.~A.~R.~Ade {\it et al.} [Planck Collaboration],
  Astron.\ Astrophys.\  {\bf 594}, A13 (2016)
  [arXiv:1502.01589 [astro-ph.CO]].

\bibitem{DarkE}
  J.~Weller and A.~M.~Lewis,
  Mon.\ Not.\ Roy.\ Astron.\ Soc.\  {\bf 346}, 987 (2003)
  [astro-ph/0307104].

\bibitem{Copeland} 
  E.~J.~Copeland, M.~Sami and S.~Tsujikawa,
  Int.\ J.\ Mod.\ Phys.\ D {\bf 15}, 1753 (2006)
  [hep-th/0603057].

\bibitem{Riess2} 
  A.~G.~Riess {\it et al.},
  Astrophys.\ J.\  {\bf 826}, no. 1, 56 (2016)
  [arXiv:1604.01424 [astro-ph.CO]];
  A.~G.~Riess {\it et al.},
  Astrophys.\ J.\  {\bf 730}, 119 (2011)
  Erratum: [Astrophys.\ J.\  {\bf 732}, 129 (2011)]
  [arXiv:1103.2976 [astro-ph.CO]].

\bibitem{Granda} 
  L.~N.~Granda and A.~Oliveros,
  Phys.\ Lett.\ B {\bf 669}, 275 (2008)
  [arXiv:0810.3149 [gr-qc]].

\bibitem{Gao} 
  C.~Gao, F.~Wu, X.~Chen and Y.~G.~Shen,
  Phys.\ Rev.\ D {\bf 79}, 043511 (2009)
  [arXiv:0712.1394 [astro-ph]];
  X.~Zhang,
  Phys.\ Rev.\ D {\bf 79}, 103509 (2009)
  [arXiv:0901.2262 [astro-ph.CO]];
  C.~J.~Feng and X.~Zhang,
  Phys.\ Lett.\ B {\bf 680}, 399 (2009)
  [arXiv:0904.0045 [gr-qc]];
  T.~F.~Fu, J.~F.~Zhang, J.~Q.~Chen and X.~Zhang,
  Eur.\ Phys.\ J.\ C {\bf 72}, 1932 (2012)
  [arXiv:1112.2350 [astro-ph.CO]].

\bibitem{SergiodelCampo} 
  S.~del Campo, J.~C.~Fabris, R.~Herrera and W.~Zimdahl,
  Phys.\ Rev.\ D {\bf 83}, 123006 (2011)
  [arXiv:1103.3441 [astro-ph.CO]].

\bibitem{Lepe} 
  S.~Lepe and F.~Pena,
  Eur.\ Phys.\ J.\ C {\bf 69}, 575 (2010)
  [arXiv:1005.2180 [hep-th]].

\bibitem{Fabiola} 
  F.~Arevalo, P.~Cifuentes, S.~Lepe and F.~Peña,
  Astrophys.\ Space Sci.\  {\bf 352}, 899 (2014)
  [arXiv:1308.5007 [gr-qc]].

\bibitem{chimentointeracciongeneral}
  L.~P.~Chimento, M.~I.~Forte and M.~G.~Richarte,
  AIP Conf.\ Proc.\  {\bf 1471}, 39 (2012)
  [arXiv:1206.0179 [gr-qc]].

\bibitem{Maldacena} 
  J.~M.~Maldacena,
  Int.\ J.\ Theor.\ Phys.\  {\bf 38}, 1113 (1999)
  [Adv.\ Theor.\ Math.\ Phys.\  {\bf 2}, 231 (1998)]
  [hep-th/9711200];
  W.~Fischler and L.~Susskind,
  hep-th/9806039.
  R.~Bousso,
  Rev.\ Mod.\ Phys.\  {\bf 74}, 825 (2002)
  [hep-th/0203101].

\bibitem{Hooft} 
  G.~'t Hooft,
  Conf.\ Proc.\ C {\bf 930308}, 284 (1993)
  [gr-qc/9310026];
  L.~Susskind,
  J.\ Math.\ Phys.\  {\bf 36}, 6377 (1995)
  [hep-th/9409089];
  J.~D.~Bekenstein,
  Phys.\ Rev.\ D {\bf 49}, 1912 (1994)
  [gr-qc/9307035].

\bibitem{Cohen} 
  A.~G.~Cohen, D.~B.~Kaplan and A.~E.~Nelson,
  Phys.\ Rev.\ Lett.\  {\bf 82}, 4971 (1999)
  [hep-th/9803132].

\bibitem{Li} 
  M.~Li,
  Phys.\ Lett.\ B {\bf 603}, 1 (2004)
  [hep-th/0403127].


\bibitem{Mathew} 
  T.~K.~Mathew, J.~Suresh and D.~Divakaran,
  Int.\ J.\ Mod.\ Phys.\ D {\bf 22}, 1350056 (2013)
  [arXiv:1207.5886 [astro-ph.CO]];
  P.~Pankunni and T.~K.~Mathew,
  Int.\ J.\ Mod.\ Phys.\ D {\bf 23}, 1450024 (2014)
  [arXiv:1309.3136 [astro-ph.CO]].

\bibitem{Ryden} B. Ryden, {\it{Introduction to Cosmology}}, Ohio State University Press, (2006).

\bibitem{interaccionconterminocomomateria} 
  S.~Chattopadhyay and A.~Pasqua,
  Indian J.\ Phys.\  {\bf 87}, 1053 (2013).


\bibitem{modeloscosmologicosCataldo} 
  M.~Cataldo, F.~Arevalo and P.~Minning,
  JCAP {\bf 1002}, 024 (2010)
  [arXiv:1002.3415 [astro-ph.CO]].


\bibitem{AICprofAntonella} 
  F.~Arevalo, A.~Cid and J.~Moya,
  Eur.\ Phys.\ J.\ C {\bf 77}, no. 8, 565 (2017)
  [arXiv:1610.09330 [astro-ph.CO]].
 

\bibitem{Bayesian} 
  A.~Cid, B.~Santos, C.~Pigozzo, T.~Ferreira and J.~Alcaniz,
  arXiv:1805.02107 [astro-ph.CO].

\end{thebibliography}
\end{document}